\documentclass[conference]{IEEEtran}
\IEEEoverridecommandlockouts

\usepackage{cite}
\usepackage{amsmath,amssymb,amsfonts}
\usepackage{algorithmic}
\usepackage{graphicx}
\usepackage{textcomp}
\usepackage{xcolor}
\usepackage{booktabs}
\usepackage{float}
\usepackage{subcaption}
\usepackage{eso-pic}      

\usepackage{hyperref}     
\newcommand{\AddCompasHeader}{%
  \begingroup
  \ifdefined\AtTextUpperRight
    \AddToShipoutPictureFG*{%
      \AtTextUpperRight{%
        \raisebox{-1.3cm}{%
          \hspace*{-6.5cm}%
          \fbox{\parbox{12cm}{\raggedleft\footnotesize
            IEEE 2nd International Conference on Computing, Applications and Systems (COMPAS 2025)\\
            23--24 October 2025, Kushtia, Bangladesh
          }}%
        }%
      }%
    }%
  \fi
  \endgroup
}

\newcommand{\compascopyright}{979-8-3315-5525-2/25/\$31.00~\copyright~2025 IEEE}
\IEEEpubid{\makebox[\columnwidth]{\compascopyright\hfill}%
\hspace{\columnsep}\makebox[\columnwidth]{}}

\begin{document}

\title{Emotion Detection in Speech Using Lightweight and Transformer-Based Models: A Comparative and Ablation Study}

\author{
\IEEEauthorblockN{1\textsuperscript{st} Lucky Onyekwelu-Udoka}
\IEEEauthorblockA{\textit{Electrical and Computer Engineering}\\
Iowa State University, Ames, USA\\
Lucky@iastate.edu}
\and
\IEEEauthorblockN{2\textsuperscript{nd} Md Shafiqul Islam}
\IEEEauthorblockA{\textit{Electrical and Computer Engineering}\\
Iowa State University, Ames, USA\\
shafiqul@iastate.edu}
\and
\IEEEauthorblockN{3\textsuperscript{rd} Md Shahedul Hasan}
\IEEEauthorblockA{\textit{Electrical and Computer Engineering}\\
Iowa State University, Ames, USA\\
shahedul@iastate.edu}
}

\maketitle
\AddCompasHeader   
\IEEEpubidadjcol   

\begin{abstract} Emotion recognition from speech plays a vital role in the development of empathetic human-computer interaction systems. This paper presents a comparative analysis of lightweight transformer-based models, DistilHuBERT and PaSST, by classifying six core emotions from the CREMA-D dataset. We benchmark their performance against a traditional CNN-LSTM baseline model using MFCC features. DistilHuBERT demonstrates superior accuracy (70.64\%) and F1 score (70.36\%) while maintaining an exceptionally small model size (0.02 MB), outperforming both PaSST and the baseline. Furthermore, we conducted an ablation study on three variants of the PaSST, Linear, MLP, and Attentive Pooling heads, to understand the effect of classification head architecture on model performance. Our results indicate that PaSST with an MLP head yields the best performance among its variants but still falls short of DistilHuBERT. Among the emotion classes, \textit{ angry} is consistently the most accurately detected, while \textit{disgust} remains the most challenging. These findings suggest that lightweight transformers like DistilHuBERT offer a compelling solution for real-time speech emotion recognition on edge devices. The code is available at: \url{https://github.com/luckymaduabuchi/Emotion-detection-}. 
\end{abstract}

\section{Introduction}
    Emotion detection from speech has become an increasingly vital area of research, with applications spanning intelligent virtual assistants, affective computing, mental health monitoring, and immersive virtual environments \cite{el2011survey, schuller2018speech}. As human-computer interactions become more natural and personalized, the demand for systems capable of interpreting emotional signals in real time has increased. Emotion-aware systems enable machines to respond empathetically to users, adjust responses based on sentiment, and improve user experience through customized feedback mechanisms \cite{akccay2020speech}. In domains such as telemarketing, adaptive education, and therapeutic interventions, emotion detection empowers analytical tools that optimize engagement and emotional relevance \cite{latif2020survey}.
    Emotion signals are typically conveyed through three modalities: facial expressions, physiological signals, and vocal audio. Among these, vocal audio presents both a rich source of emotional information and a challenging recognition problem. Compared to image-based facial cues, speech provides more dynamic, personalized, and nuanced emotional content \cite{kim2013deep}. However, the complexity of speech, driven by factors such as tone, prosody, speaker identity, and conversational context, makes emotion recognition from audio an open-ended machine learning challenge. The feature extraction process, the choice of representation (MFCC versus spectrograms), and the model architecture significantly influence the system’s ability to reliably decode emotions \cite{lee2021comprehensive}.
    Earlier approaches relied heavily on statistical methods such as Gaussian Mixture Models with Universal Background Models (GMM-UBM) and hybrid classifiers such as GMM-DNN~\cite{vydana2015,shahin2019}. 
    Although effective for constrained settings, these models struggled to scale to large, diverse datasets due to limitations in sequential modeling and robustness to noise. 
    The introduction of deep learning models, especially CNN-LSTM architectures using hand-crafted features such as MFCCs, marked a turning point, improving both performance and temporal modeling. 
    Ouyang et al.~\cite{ouyang2024} demonstrated such improvements using a CNN-LSTM pipeline on MFCC-transformed speech data, achieving an accuracy of 61.07\%.
    More recently, transformer-based architectures have revolutionized speech representation learning. 
    Self-supervised models such as DistilHuBERT leverage layer-wise knowledge distillation to offer high accuracy with minimal computational overhead~\cite{wang2021distilhubert}. 
    In parallel, PaSST~\cite{zeineldeen2021passt}, designed for efficient audio classification, introduces spectrogram patching and patchout techniques to generalize effectively. 
In this study, we present a comparative analysis of DistilHuBERT, PaSST, and a CNN-LSTM baseline for the classification of speech emotions using the CREMA-D dataset. Furthermore, we conduct an ablation study on PaSST configurations to understand how architectural variations (linear and attention vs. MLP heads) and how raw audio vs. spectrogram input impact performance.

\section*{Related Work}
Early work in speech emotion recognition (SER) leveraged primarily hand-crafted acoustic features such as Mel frequency cepstral coefficients (MFCC) or spectrograms, which were input to deep learning architectures like Convolutional Neural Networks (CNN) or Recurrent Neural Networks (RNN) \cite{vydana2015, shahin2019}. CNNs, in particular, were effective in capturing localized spectral-temporal patterns. Fayek et al. used deep CNNs to classify emotions from spectrogram images, achieving a precision of around 60\% on the SAVEE corpus \cite{fayek2017}. Later, more advanced CNN architectures improved this, especially when combined with global pooling layers or augmented feature sets, achieving up to 70\% accuracy in datasets like RAVDESS \cite{ravdess2018}.
RNNs, especially long-short-term memory networks (LSTM), were also widely adopted due to their ability to model temporal dependencies in sequential speech data \cite{mirsamadi2017}. LSTM networks operating on MFCC sequences demonstrated performance comparable to CNNs, particularly in modeling prosodic features such as rhythm and pitch contours. Some architectures combined CNN and RNN modules, CNNs to extract spatial features, and RNNs to model temporal dynamics, achieving enhanced performance \cite{trigeorgis2016}. For example, Trigeorgis \textit{et al.} proposed an end-to-end convolutional recurrent network learning directly from raw waveforms \cite{trigeorgis2016}.

The attention mechanisms further improved these models. By integrating attention layers on top of LSTMs or CNNs, models could focus on the most emotionally salient parts of an utterance. Mountzouris \textit{et al.} achieved more than 74\% accuracy on SAVEE and 77\% on RAVDESS using CNN-attention hybrids \cite{mountzouris2022}. However, despite these gains, CNN and RNN-based models often struggled with generalization due to limited dataset sizes and speaker variability, prompting a shift toward self-supervised and pre-trained models \cite{pepino2021emotion, tripathi2020self}.

Transformer models have more recently become prominent in SER due to their ability to model long-range dependencies and benefit from large-scale pretraining. Among them, wav2vec 2.0 is a leading self-supervised model trained in raw audio using contrastive learning \cite{pepino2021emotion}. It consists of a convolutional encoder followed by a Transformer that captures contextual dependencies. Pepino \textit{et al.} demonstrated that fine-tuned wav2vec 2.0 models outperform previous CNN/LSTM models, achieving up to 73\% accuracy on IEMOCAP \cite{pepino2021emotion}.

HuBERT (Hidden Unit BERT), another Transformer-based model, differs by using masked prediction of cluster-based units derived from acoustic features \cite{hsu2021hubert}. Fine-tuned HuBERT models have shown even higher SER accuracy, reaching up to 79.6\% on IEMOCAP and exhibiting strong performance on individual emotions such as anger and fear.

To reduce computational complexity, DistilHuBERT was proposed as a distilled version of HuBERT \cite{wang2021distilhubert}. It compresses the model by 75\% and accelerates inference while maintaining competitive performance, making it ideal for real-time applications.

Another line of work uses Transformers on spectrogram images. The Audio Spectrogram Transformer (AST) and its efficient variant PaSST (Patchout Spectrogram Transformer) apply the Vision Transformer (ViT) framework to audio spectrograms \cite{zeineldeen2021passt}. PaSST incorporates patchout regularization, which randomly drops time/frequency patches during training, reducing memory usage and acting as an augmentation. These models have achieved strong results on AudioSet and have been adapted for SER tasks.

The evolution from CNN/RNN models to Transformer-based architectures has significantly improved SER accuracy, robustness, and efficiency. Transformer models benefit from self-attention, allowing them to capture both global and fine-grained prosodic features. Pre-training on large speech corpora enables better generalization even on smaller SER datasets.

While large Transformers like wav2vec 2.0 and HuBERT deliver superior performance, they are computationally intensive. Models like DistilHuBERT and PaSST strike a balance between accuracy and efficiency, making them practical for deployment. 

\section*{Problem Definition}

Given a raw audio signal \( \mathbf{x}(t) \), the task of Speech Emotion Recognition (SER) is to classify the signal into one of \( K \) discrete emotion classes:
\[
\mathcal{Y} = \{\text{happy}, \text{sad}, \text{angry}, \text{fear}, \text{disgust}, \text{neutral}\}
\]

Let \( \mathbf{x} \in \mathbb{R}^T \) denote a speech waveform of duration \( T \), and let \( f_\theta: \mathbb{R}^T \rightarrow \mathbb{R}^K \) be a parameterized model (e.g., a Transformer-based or CNN-based architecture). The goal is to learn the mapping:
\[
\hat{\mathbf{y}} = f_\theta(\mathbf{x})
\]
where \( \hat{\mathbf{y}} \in \mathbb{R}^K \) is the predicted probability distribution over the emotion classes, and the final predicted label is:
\[
\hat{y} = \arg\max_i \hat{y}_i
\]

Training is performed by minimizing the categorical cross-entropy loss between the predicted distribution \( \hat{\mathbf{y}} \) and the ground truth label \( \mathbf{y} \in \{0, 1\}^K \):
\[
\mathcal{L}(\mathbf{y}, \hat{\mathbf{y}}) = -\sum_{i=1}^{K} y_i \log(\hat{y}_i)
\]

This paper addresses the problem of identifying the most accurate and efficient model architecture for real-time SER under consistent training and evaluation conditions. Specifically, our goal is to:

\begin{itemize}
    \item Compare the performance of a lightweight self-supervised model (DistilHuBERT), a spectrogram-based Transformer (PaSST), and a CNN-LSTM baseline.
    \item Evaluate the effect of different classification heads in PaSST: Linear, MLP, and Attentive Pooling.
    \item Identify which model offers the best trade-off between accuracy, inference time, and memory efficiency in the CREMA-D dataset.
\end{itemize}

By benchmarking these models and configurations, we seek to provide insights into optimal architectures for practical deployment of SER systems on resource-constrained devices.
\section*{Methodology}
 The model choices reflect different levels of abstraction and learning paradigms: CNN-LSTM illustrates sequential modeling from engineered features (MFCC), DistilHuBERT demonstrates self-supervised representation learning directly from waveforms, and PaSST showcases transformer-based architectures for image-like inputs such as spectrograms. These models provide a balanced overview of both classical and contemporary approaches.

\subsection*{Dataset and Preprocessing}
The Crowd-sourced Multimodal Emotional Actors Dataset (CREMA-D) is used in this study. It contains 7,442 audio clips from 91 actors who speak 12 sentences in six basic emotional states. Anger, Disgust, Fear, Happy, Neutral, and Sad. The data set provides diverse speakers in terms of age, gender, and ethnicity, making it suitable for training robust emotion recognition models\cite{cao2014crema}.

Each audio file in the CREMA-D dataset is loaded at a target sampling rate of 16 kHz and clipped or padded to a maximum duration of 10 seconds. During training, several forms of data augmentation are applied to improve model generalization. These include random gain adjustment, where a gain between -6 dB and +6 dB is applied, additive Gaussian noise to simulate background interference, pitch shifting simulated through resampling to slightly higher or lower sampling rates, and then converting back to 16 kHz, and random time shifting by circularly rolling the waveform forward or backward in time. All audio waveforms are normalized and returned along with their categorical emotion label for supervised learning.

\subsection*{Models and Implementation}
The baseline model used in this study is a CNN-LSTM hybrid architecture that operates on Mel frequency cepstral coefficients (MFCC) as input features. Reproduced from Ouyang et al.~\cite{ouyang2024}, the model consists of four convolutional layers followed by three LSTM layers and a fully connected classification head, achieving an accuracy of 61.07\% on the CREMA-D dataset.
The 2D convolutional stack captures local spectral and temporal features, while the bidirectional LSTM layers model sequential dependencies in the speech signal. This combination allows the network to learn both spatial and temporal patterns, making it a strong and well-established classical baseline for speech emotion recognition tasks.

\subsubsection*{DistilHuBERT}
DistilHuBERT is a lightweight, distilled version of the HuBERT speech model. It comprises a convolutional feature extractor and a 2-layer Transformer encoder, distilled from a 12-layer HuBERT model using layer-wise knowledge distillation \cite{chang2022distilhubert}. The model takes raw waveforms as input and output embeddings representing phonetic and prosodic information. For classification, a linear head is attached to the CLS token representation or the mean of the hidden states.

\begin{figure}[t]
    \centering
    \includegraphics[width=0.7\linewidth]{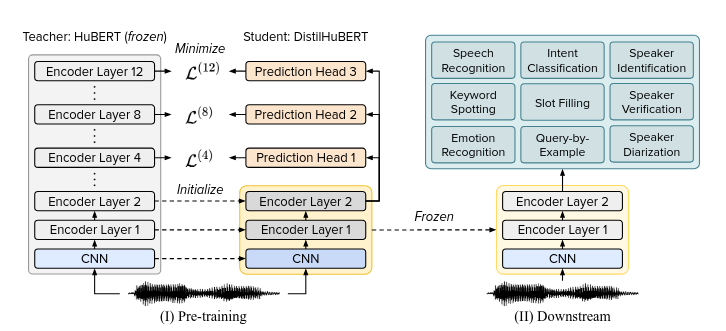}
    \caption{DistilHuBERT architecture overview.}
    \label{fig:distilhubert}
\end{figure}

\subsubsection*{Patchout Spectrogram Transformer}
\begin{figure}[H]
    \centering
    \includegraphics[width=0.5\linewidth]{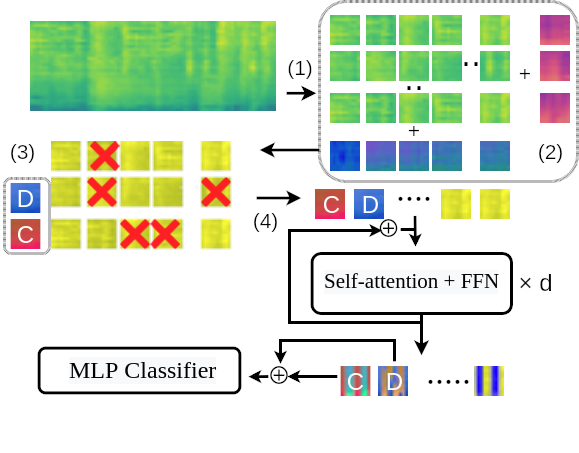}
    \caption{PaSST}
    \label{fig:enter-label}
\end{figure}

The Patchout Spectrogram Transformer (PaSST) adapts the Vision Transformer (ViT) framework to audio spectrograms. Input spectrograms are divided into fixed-size patches that are flattened and projected into an embedding space. Two positional embeddings-time and frequency are added, and the sequence is passed through 12 transformer blocks with multihead self-attention and MLP layers.
PaSST introduces a regularization technique called \textit{patchout}, which randomly drops time-frequency patches during training, acting as both a regularizer and an augmentation. For classification, both linear and MLP heads are tested\cite{koutini2022passt}.

\subsubsection*{Ablation Study Setup}

To assess how architectural variations affect the performance of the PaSST model, an ablation study is conducted by experimenting with different classification heads and training configurations. All experiments use the pre-trained \texttt{passt\_s\_swa\_p16\_128\_ap476} backbone with patchout enabled for regularization. The specific configurations include:

\textbf{Linear Head}: This is the PaSST configuration, where classification is performed by applying a single linear transformation to the output of the [CLS] token. Let \( \mathbf{h}_{\text{cls}} \in \mathbb{R}^{d} \) be the CLS embedding; then the logits are computed as:
\[
\mathbf{z} = \mathbf{W} \mathbf{h}_{\text{cls}} + \mathbf{b}, \quad \mathbf{z} \in \mathbb{R}^K
\]
Only the final transformer block (block 11) and the classifier layer are frozen for fine-tuning.

\textbf{MLP Head}: A two-layer feedforward network is applied to \( \mathbf{h}_{\text{cls}} \), consisting of LayerNorm, ReLU, Dropout, and a linear output layer. The formulation is:
\[
\mathbf{h}_1 = \text{ReLU}(\mathbf{W}_1 \cdot \text{LayerNorm}(\mathbf{h}_{\text{cls}}) + \mathbf{b}_1)
\]
\[
\mathbf{z} = \mathbf{W}_2 \cdot \text{Dropout}(\mathbf{h}_1) + \mathbf{b}_2
\]
Here, \( \mathbf{W}_1 \in \mathbb{R}^{256 \times 768} \), \( \mathbf{W}_2 \in \mathbb{R}^{K \times 256} \). The MLP head and the last two transformer blocks (blocks 10 and 11) are unfrozen during fine-tuning.

\textbf{Attentive Pooling Head}: Instead of using the CLS token, this configuration aggregates all token embeddings \( \mathbf{H} = [\mathbf{h}_1, \dots, \mathbf{h}_T] \in \mathbb{R}^{T \times d} \) using attention weights:
\[
\alpha_t = \frac{\exp(\mathbf{w}_2^\top \tanh(\mathbf{W}_1 \mathbf{h}_t))}{\sum_{j=1}^T \exp(\mathbf{w}_2^\top \tanh(\mathbf{W}_1 \mathbf{h}_j))}
\]
\[
\boldsymbol{\mu} = \sum_{t=1}^T \alpha_t \mathbf{h}_t, \quad \boldsymbol{\sigma} = \sqrt{\sum_{t=1}^T \alpha_t (\mathbf{h}_t - \boldsymbol{\mu})^2}
\]
\[
\mathbf{z} = \mathbf{W} [\boldsymbol{\mu}; \boldsymbol{\sigma}] + \mathbf{b}
\]
The attention module and the last two transformer blocks are trainable in this setup.

\begin{figure}
    \centering
    \includegraphics[width=0.35\linewidth]{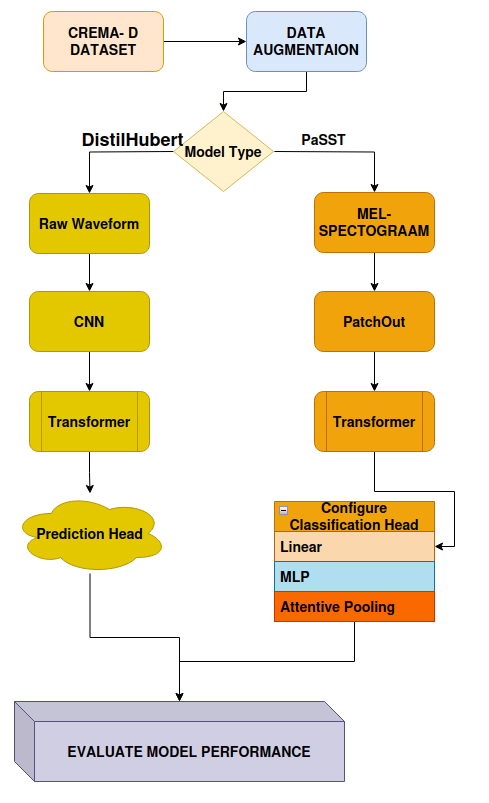}
     \caption{Overall training pipeline for DistilHuBERT and PaSST models on the CREMA-D dataset.}
    \label{fig:training_pipeline}
\end{figure}

Figure~\ref{fig:training_pipeline} shows the training and evaluation pipeline for speech emotion recognition using DistilHuBERT and PaSST. The process begins with the CREMA-D dataset, where the audio samples undergo data augmentation. For DistilHuBERT, raw waveforms are processed by a CNN and Transformer backbone to extract contextual embeddings. In contrast, PaSST transforms audio into Mel spectrograms, applies patchout regularization, and forwards the result through a transformer. PaSST models use configurable classification heads: linear, MLP, or attentive pooling before proceeding to evaluation. Performance metrics are computed to assess the comparative effectiveness of both architectures.

\subsection*{Training and Evaluation}
All models were trained with identical protocols for a fair comparison. We used a speaker-independent split of CREMA-D (70\% training, 15\% validation, 15\% test; no speaker overlap). This split strategy prevents data leakage and supports fair generalization evaluation.

Training was carried out for up to 30 epochs with early stopping in validation accuracy (patience = 5). The optimizer was Adam with $\beta_1=0.9$, $\beta_2=0.999$; the initial learning rate was $1\times10^{-4}$ with cosine decay and no warmup; batch size was 16; and the loss was macro-averaged cross-entropy calculated over classes. The experiments were carried out on a single NVIDIA GPU (RTX 4090).

Model performance was evaluated using accuracy, precision, recall, and the F1 score. In addition, confusion matrices were generated to analyze class-wise recognition performance. To assess the feasibility of deployment, we also reported inference time per sample (milliseconds) and the total size of each model (megabytes).

\subsection*{Results and Comparative Analysis}
Table~\ref{tab:model_comparison}  presents the overall performance metrics for all models evaluated in the CREMA-D test set. DistilHuBERT achieved the highest overall accuracy (70.64\%) and the F1 score (70.36\%), while requiring only 0.02 MB in size and maintaining competitive inference time, making it the most efficient and accurate among the models evaluated. 
Among the PaSST variants, the MLP head model performed the best accuracy (54.07\%), closely followed by the attentive pooling and linear head configurations. All variants shared the same input representation, Mel spectrograms, but differed in how the extracted Transformer features were aggregated and classified. In particular, the attentive pooling head, which summarizes temporal token features using learned statistical attention, outperformed the simpler linear projection. This challenges the notion that basic classification heads suffice when using spectrogram-based inputs, instead showing that expressive heads can extract more emotionally salient information. 
In particular, not all transformer-based models outperformed traditional architectures. The CNN-LSTM baseline achieved 61. 07\% precision, significantly surpassing all PaSST configurations. This result shows the strength of RNN-based temporal modeling and the value of simpler architectures, especially when dealing with moderately sized datasets such as CREMA-D. 

\begin{table*}[h]
\centering
\begin{tabular}{lcccccc}
\toprule
\textbf{Model} & Accuracy & F1-score & Precision & Recall & Inf. Time (ms) & Size (MB) \\
\midrule
DistilHuBERT         & \textbf{70.64\%} & \textbf{70.36\%} & 71.67\% & 70.64\% & 21.4  & \textbf{0.02} \\
PaSST (MLP)          & 54.07\% & 53.82\% & 54.28\% & 54.07\% & 19.0  & 342.21 \\
PaSST (Raw)          & 52.46\% & 52.05\% & 52.70\% & 52.46\% & 19.0  & 341.00 \\
PaSST (Linear)       & 52.15\% & 51.29\% & 51.98\% & 51.75\% & \textbf{18.5} & 341.41 \\
CNN-LSTM (Baseline)  & 61.07\% & ---     & ---     & ---     & ---   & ---    \\
\bottomrule
\end{tabular}
\caption{Model comparison on CREMA-D dataset}
\label{tab:model_comparison}
\end{table*}

\begin{table}[H]
\centering
\resizebox{0.5\linewidth}{!}{ 
\begin{tabular}{lccc}
\toprule
\textbf{Emotion} & DistilHuBERT & PaSST-MLP & CNN-LSTM \\
\midrule
Angry    & \textbf{86.91\%} & 68.22\% & 75.31\% \\
Neutral  & 71.72\% & 66.49\% & \textbf{71.70\%} \\
Happy    & 63.35\% & \textbf{59.17\%} & 61.18\% \\
Sad      & 54.45\% & 54.55\% & 56.70\% \\
Fear     & 67.37\% & \textbf{60.47\%} & 59.04\% \\
Disgust  & 40.31\% & \textbf{43.46\%} & 38.33\% \\
\bottomrule
\end{tabular}
}
\caption{Per-emotion classification accuracy}
\label{tab:emotion}
\end{table}

\begin{table*}
\centering
\begin{tabular}{lccp{9cm}}  
\toprule
\textbf{Configuration} & \textbf{Accuracy} & \textbf{F1-score} & \textbf{Notes} \\
\midrule
Linear Head            & 52.15\% & 51.29\% & Minimal design using a single linear projection of the [CLS] token without additional non-linearity or pooling. Yields the lowest performance. \\
MLP Head               & \textbf{54.07\%} & \textbf{53.82\%} & A two-layer feedforward network with ReLU activation and optional dropout. Applies LayerNorm. Provides the best results in classification. \\
Attentive Pooling Head & 52.46\% & 52.05\% & Replaces [CLS] token with attention-weighted aggregation over all frame tokens. Captures contextual relevance better than Linear but underperforms MLP. \\
\bottomrule
\end{tabular}
\caption{Ablation study of different classification heads in PaSST for emotion recognition.}
\label{tab:ablation}
\end{table*}

\begin{figure}[t]
    \centering
    \begin{subfigure}[t]{0.4\linewidth}
        \centering
        \includegraphics[width=\linewidth]{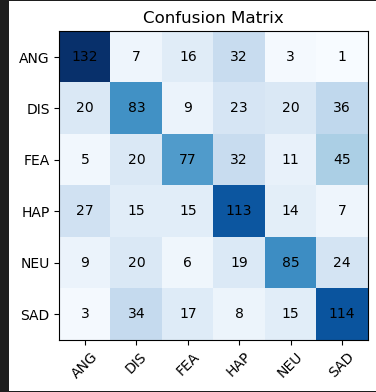}
        \caption{PaSST-MLP}
        \label{fig:conf_passt_mlp}
    \end{subfigure}
    \hfill
    \begin{subfigure}[t]{0.48\linewidth}
        \centering
        \includegraphics[width=\linewidth]{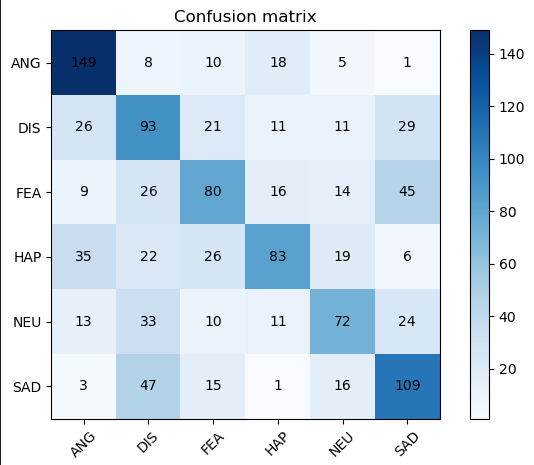}
        \caption{PaSST-Attention}
        \label{fig:conf_passt_attn}
    \end{subfigure}

    \caption{Confusion matrices of PaSST-MLP and PaSST-Attention showing per-emotion classification performance on the CREMA-D dataset.}
    \label{fig:conf_part2}
\end{figure}

\begin{figure}[t]
    \centering
    \begin{subfigure}[t]{0.48\linewidth}
        \centering
        \includegraphics[width=\linewidth]{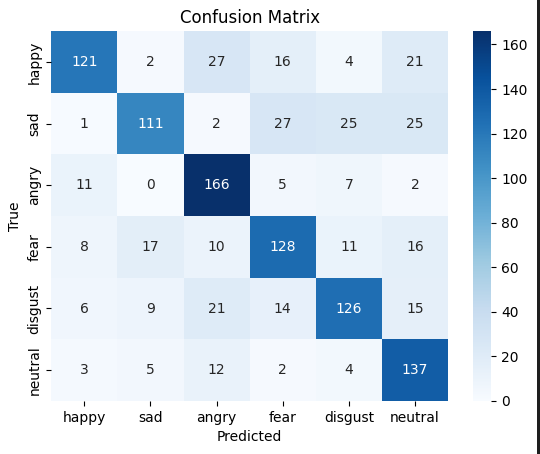}
        \caption{DistilHuBERT}
        \label{fig:conf_distil}
    \end{subfigure}
    \hfill
    \begin{subfigure}[t]{0.4\linewidth}
        \centering
        \includegraphics[width=\linewidth]{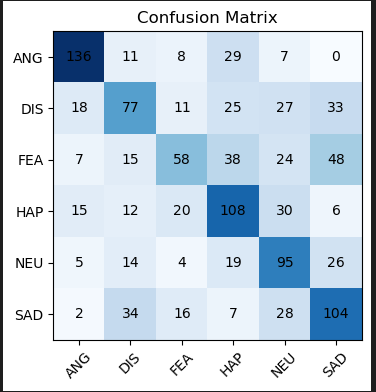}
        \caption{PaSST-Linear}
        \label{fig:conf_passt_linear}
    \end{subfigure}

    \caption{Confusion matrices of DistilHuBERT and PaSST-Linear showing per-emotion classification performance on the CREMA-D dataset.}
    \label{fig:conf_part1}
\end{figure}

Table~\ref{tab:emotion} summarizes the accuracy of the classification per emotion. DistilHuBERT clearly excelled in recognizing high arousal emotions such as \textit{Angry} (86.91\%) and \textit{Fear} (67.37\%), showcasing its capacity to capture expressive variations in speech. For subtler emotions like \textit{Neutral} and \textit{Sad}, both CNN-LSTM and PaSST-MLP showed competitive performance, reflecting their potential to model more nuanced or flat affective tones.
The emotion that performed the worst in all models was \textit{Disgust}, likely due to its low frequency of occurrence and ambiguous acoustic features. Surprisingly, the PaSST-MLP variant achieved a slight edge here (43.46\%), suggesting that spectrogram-based attention may still capture isolated emotional cues better in rare categories.

\subsection*{Visual Comparison of Model Interpretations}
To further clarify how models process emotion-labeled audio, consider an example from the CREMA-D data set labeled 'Angry.' This audio sample undergoes data augmentation, including gain adjustment and pitch shift, simulating real-world recording variations. For the CNN-LSTM model, the sample is converted into MFCC features and passed through convolutional layers that capture local spectral patterns and LSTM layers that model temporal dynamics. DistilHuBERT processes the raw waveform directly, extracting contextual embeddings through its convolutional and transformer layers. PaSST, on the other hand, converts the audio into a Mel spectrogram and processes it via transformer blocks using patch-based attention.

\begin{figure}[H]
\centering
\includegraphics[width=0.3\linewidth]{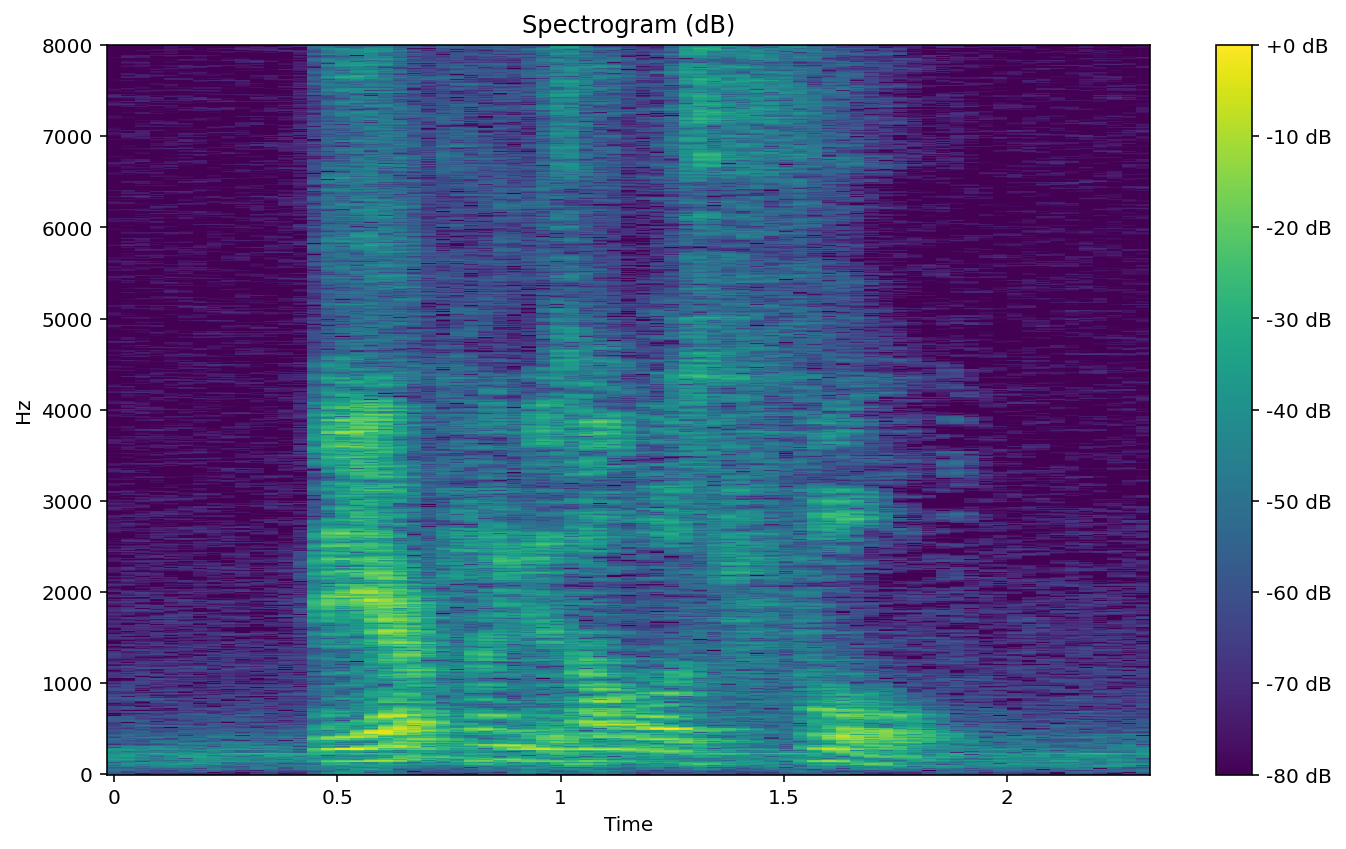}
\caption{Mel spectrogram of an 'Angry' utterance from the CREMA-D dataset. Energy is concentrated in the lower frequencies with noticeable bursts around 1–2 kHz and 3–4 kHz.}
\label{fig:toy_spectrogram}
\end{figure}

In this specific case, DistilHuBERT correctly classifies the sample as 'Angry' with confidence 87\%, benefiting from its input of raw waveform and the ability to model prosodic cues such as pitch and tone. The model misclassifies the same sample as 'Happy', probably because of overlapping high-frequency energy in the spectrograms of both emotions. \\
Figure~\ref{fig:toy_spectrogram} shows the input of the spectrogram used by PaSST. High-intensity regions (yellow-green) appear in the lower and mid frequency bands, typical of emotionally charged speech such as anger, which tends to exhibit higher pitch variation and energy bursts. PaSST processes this patchwise, potentially missing subtle temporal cues that DistilHuBERT captures from the raw waveform.

In all of our settings, DistilHuBERT outperforms the PaSST variants (Tables~\ref{tab:model_comparison}, \ref{tab:emotion}). We hypothesize three contributing factors. First, DistilHuBERT benefits from self-supervised pre-training directly on raw waveforms, which preserves fine-grained prosodic cues (pitch contours, micropauses) that are critical for SER and can be partially smoothed by spectrogram patching. Second, PaSST’s patch-based tokenization and patchout regularization, while effective for broad audio tagging, may discard short-lived emotional microevents (e.g., bursts, glottal onsets) that matter in smaller SER datasets. Third, the data scale: with a moderate-sized CREMA-D, shallow classification heads plus limited fine-tuning may be insufficient to fully adapt a large spectrogram transformer. 
\section*{Ablation Study on PaSST Variants}

The PaSST architecture offers flexibility in how final classification is performed, allowing interchangeable classification heads. This ablation study evaluates three such configurations, Linear, MLP, and Attentive Pooling, under identical training conditions to isolate the impact of the classification head on model performance. In particular, all variants share the same input representation: Mel spectrograms. The difference lies solely in how the final emotion prediction is computed from the Transformer output.

As shown in Table~\ref{tab:ablation}, the Linear Head configuration achieved the lowest performance, with an accuracy of 52.15\% and an F1 score of 51.29\%. This configuration directly maps the \texttt{[CLS]} token embedding to the emotion classes using a single linear layer. Although computationally efficient, its limited expressive power may constrain its ability to capture complex emotional nuances.

The MLP Head achieved the highest accuracy at 54.07\% and an F1 score of 53.82\%. The added depth and non-linearity enable richer abstraction of features, demonstrating the effectiveness of moderately complex heads for emotion recognition.
The Attention Group Head replaced the default \texttt{[CLS]} token with a statistical grouping mechanism applied to all temporal tokens. This combination computes a weighted mean and standard deviation of token characteristics, with attention weights learned during training. Although it performed slightly better than the linear head (52.46\% accuracy, 52.05\% F1), it still lagged behind the MLP configuration. This suggests that attention-based statistics help summarize temporal features, but may not be sufficient without additional nonlinear transformations.

These results indicate that head design plays a critical role in SER performance. Even when the backbone of the transformer and the input of the spectrogram remain constant, the classification head capacity significantly influences the model’s ability to discriminate emotional states.

\section*{Conclusion and Future Work}

Although PaSST performed poorly relative to DistilHuBERT, this study offers a modular and interpretable architecture with tunable classification heads. Specifically, the MLP head provided the best results, indicating that shallow nonlinear transformations can help extract more discriminative features from transformer output. 
Raw waveform models such as DistilHuBERT are better suited to capture prosodic and temporal features, while spectrogram-based models like PaSST require more careful architectural tuning to compete.

Future work may explore the integration of multimodal signals, such as visual and physiological cues, to improve emotion recognition under ambiguous or low-quality audio conditions. Furthermore, extending PaSST pretraining to emotion-rich datasets and incorporating emotion-aware objectives during fine-tuning could help bridge the performance gap with DistilHuBERT. 

\bibliographystyle{IEEEtran}
\bibliography{refs}

\end{document}